\documentclass[article]{IEEEtran}
\usepackage{amsmath,amssymb,color,graphicx,caption,subcaption,comment,tikz,url} 
\usepackage{pgfplots,epstopdf}
\usepackage{verbatim}
\usetikzlibrary{decorations.markings}
\usepackage{cite}
\usepackage{enumerate}
\usepackage{bbm}
\usepgfplotslibrary{units}
\usetikzlibrary{spy,backgrounds}
\usepackage{pgfplotstable}
\newtheorem{theorem}{Theorem}

\newtheorem{lemma}{Lemma}

\newtheorem{definition}{Definition}

\newtheorem{remark}{Remark}

\newcommand{\mbf}[1]{\mathbf{#1}}
\newcommand{\set}[1]{\mathcal{#1}}

\newcommand{\E}{\mathbb{E}}

\newcommand{\M}{M}
\newcommand{\D}{\mbf{D}}
\renewcommand{\d}{\mbf{d}}

\newcommand{\R}{R}

\newcommand{\V}{\mathbb{V}}
\newcommand{\Qell}{\mathcal{Q}_\ell^{\textnormal{dist}}}
\newcommand{\Qkone}{\mathcal{Q}_{k-1}^{\textnormal{dist}}}
\newcommand{\Qrepell}{\mathcal{Q}_\ell^{\textnormal{rep}}}
\newcommand{\Qrep}{\mathcal{Q}^{\textnormal{rep}}}
\newcommand{\RMworst}{R_{\textnormal{worst}}^\star(\M)}
\newcommand{\RMavg}{R_{\textnormal{avg}}^\star(\M)}
\newcommand{\Rlowerworst}{R_{\textnormal{worst}}^{\textnormal{low}}}
\newcommand{\Rloweravg}{R_{\textnormal{avg}}^{\textnormal{low}}}
\newcommand{\RlowerworstUnd}{\underline{R}_{\textnormal{worst}}}
\newcommand{\RloweravgUnd}{\underline{R}_{\textnormal{avg}}}

\allowdisplaybreaks[4] 
\graphicspath{{.}{./Figures/}} 

\IEEEoverridecommandlockouts

\begin{document}
\title{Improved Converses and Gap Results for Coded Caching}
\author{Chien-Yi Wang, Shirin Saeedi Bidokhti~\IEEEmembership{Member,~IEEE}, and Mich\`{e}le Wigger~\IEEEmembership{Senior Member,~IEEE},
\thanks{C.-Y. Wang is with MediaTek Inc., Hsinchu 30078, Taiwan. He has previously been with LTCI, Telecom ParisTech, 75013 Paris, France. (e-mail: chien-yi.wang@mediatek.com)
	S.~Saeedi~Bidokhti is with the Department of Electrical and Systems Engineering at the University of Pennsylvania, USA. (e-mail:  saeedi@seas.upenn.edu.)  S.~Saeedi~Bidokhti was supported by the Swiss National Science Foundation fellowship no. 158487. M. Wigger is with LTCI, Telecom ParisTech,  75013 Paris, France (e-mail: michele.wigger@telecom-paristech.fr).} 
\thanks{The material in this paper  was presented at the \emph{2017 IEEE International Symposium on Information Theory}, Aachen, Germany. }}

\maketitle

\begin{abstract}
Improved lower bounds are derived on the average and  worst-case rate-memory tradeoffs of the Maddah-Ali\&Niesen coded caching scenario. For any number of users
and files and for arbitrary cache sizes, the multiplicative gap
between the exact rate-memory tradeoff and the new lower bound
is shown to be less than 2.315 in the worst-case scenario and  2.507
in the average-case scenario. 
\end{abstract}



\section{Introduction}\label{sec:definition}


We consider the canonical coded caching scenario by Maddah-Ali and Niesen \cite{maddahali_niesen_2014-1} with a single transmitter and $K$ receivers, where each receiver is equipped with a cache memory of equal size (see in Figure~\ref{fig:source_model}). Communication takes place in two phases: a \emph{caching phase} and a subsequent \emph{delivery phase}. In the {caching phase}, the transmitter stores contents (arbitrary functions of files) at the receivers' cache memories. In the {delivery phase}, each receiver makes a demand and the transmitter accordingly conveys the desired files to each of the receivers. The main challenge in this configuration is that during the caching phase it is not known which receiver demands  which specific file from the library.  The cache contents thus need to be designed so as to be useful for many possible demands. 

Traditional caching systems store  a portion of the most popular files in each and every cache memory. This allows the receivers to retrieve these files locally without burdening the common communication link from the transmitter to the receivers.  Recently in \cite{maddahali_niesen_2014-1}, it was shown that much larger gains, so called \emph{global caching gains}, are possible if   various receivers store different parts of the files in their cache memories. In this case, the  transmitter can  simultaneously serve multiple receivers during the delivery phase by sending coded data, and thus significantly reduce the 
  delivery rate (latency) of communication. 

{The main quantity of interest in this work is the \emph{rate-memory tradeoff} introduced in \cite{maddahali_niesen_2014-1}---i.e., the minimum required delivery rate, as a function of the  cache memories, so that all receivers reliably recover their demanded files. We consider both  the \emph{worst-case} rate-memory tradeoff defined in \cite{maddahali_niesen_2014-1}, which is the  common scenario in the coded-caching literature, as well as the \emph{average-case} rate-memory tradeoff  defined in \cite{niesen_maddahali_2014}. In the latter case, the rate can adapt to the receivers' demands and the rate-memory tradeoff is  defined as the average rate  over all possible demand vectors. Upper bounds on the worst-case rate-memory tradeoff  were presented for certain special cases in \cite{chenfanletaief-2014,tian-2015,wan_tuninetti_piantanida-2016,sahraeigastpar-2016-1,tianchen-2016,amirigunduz-2016,YuMA:16,gomez-2017} and  lower bounds (converse results) were presented in \cite{ghasemi_ramamoorthy, sengupta_tandon_clancy-2015,wanglimgastpar-2016}. The previously best lower and upper bounds for the worst-case rate-memory tradeoff  match up to a multiplicative gap of $4$ \cite{ghasemi_ramamoorthy}. 
The works in \cite{wan_tuninetti_piantanida-2016, YuMA:16} determined the exact rate-memory tradeoff assuming uncoded cache placements, i.e., assuming  that \emph{fractions of contents are cached}. This caching strategy is however known to be suboptimal in general.}
{Upper and lower bounds on the rate-memory tradeoff for the average-case scenario were derived in \cite{wanglimgastpar-2016, YuMA:16}. The previously best lower and upper bounds in this scenario have been shown to match up to a multiplicative gap of $4.7$ \cite{wanglimgastpar-2016}.}

In this paper we provide new lower bounds on  the rate-memory tradeoff. The new lower bounds match the  worst-case and average-case rate-memory tradeoffs up to multiplicative gaps of $2.315$ and $2.507$, respectively. More precisely, these gaps are with respect to the upper bounds on the rate-memory tradeoff under \emph{decentralized} caching in \cite{YuMA:16}. An upper bound on the  rate-memory tradeoff under decentralized caching is also an upper bound on the rate-memory tradeoff under \emph{centralized} caching  considered here, because in the decentralized caching the cache content at a given receiver has  to be chosen according to a specific distribution, whereas in centralized caching any content can be cached that satisfies the cache memory constraints. {In an independent and concurrent work, \cite{YMA:17} presents slightly improved lower bounds  for both the worst-case and average-case rate-memory tradeoffs. These bounds are within a gap of $2.00884$ from the decentralized schemes. The proof in \cite{YMA:17} is  similar to the proof here, but includes an additional averaging step over all possible labelings of receivers.}

Many other variations of the caching problem have recently been studied such as  {online caching} \cite{pedarsani_maddahali_niesen_2015};  caching with non-uniform demands \cite{niesen_maddahali_2014,ji_tulino_llorca_caire_2015,zhanglinwang-2015,HashemKaramchandaniDiggavi2014}; caching of correlated files \cite{Timo-Mar-2016-C,hassanzadeh-ISIT,hassanzadeh-ITW,hassanzadeh-ISTC,wang_lim_gastpar_2015,yanggunduz-2017,YangHassanGunduzErkip18} where \cite{Timo-Mar-2016-C} shows how Wyner's and Gac-K\"orner's common information play a key role; caching in noisy broadcast channels \cite{huang_wang_ding_yang_zhang_2015,timowigger-2015-1,saeedibidokhti_timo_wigger-2016,saeediwiggertimo-turbo,tulino_jccs,GunduzBC17,GunduzGBC17,hassanzadeherkipllorcatulino-2015,ghorbelkobayashiyang-2016-2,zhangengelmannelia-2015,zhangelia-2015,zhangelia-2016-2,zhangelia-2016,yangngokobayashi-2016,Shariatpanahi18} where coding opportunities could be exploited through joint cache-channel coding in heterogeneous networks \cite{timowigger-2015-1,saeedibidokhti_timo_wigger-2016,tulino_jccs,GunduzBC17}\footnote{{Joint cache-channel coding is used for joint source-channel coding where part of the sources are actually given by the cache contents.}}, feedback and channel state information \cite{ghorbelkobayashiyang-2016-2,zhangelia-2015,zhangengelmannelia-2015, zhangelia-2016-2}, and multiple antennas \cite{yangngokobayashi-2016,Shariatpanahi18};  caching in Gaussian interference networks \cite{maddahaliniesen-2015-1,naderializadehmaddahaliavestimehr-2016, SenguptaTandonOsvaldo}; hierarchical networks~\cite{nikhil1} and multi-server networks \cite{pooyaabolfazlhossein-2015}; and cellular networks~\cite{ShanmugamGolrezaeietal13,ntranosmaddahalicaire-2015,ugurawansezgin-2015, yicaire-2016,parksimeoneshamai-2016,tandonsimeone-2016,azarisimeonespagnolinitulino-2016,baszczyszyngiovanidis-2014}.

\section{Detailed Problem Setup}

\begin{figure}[t]
\begin{center}
\includegraphics[width=0.45\textwidth]{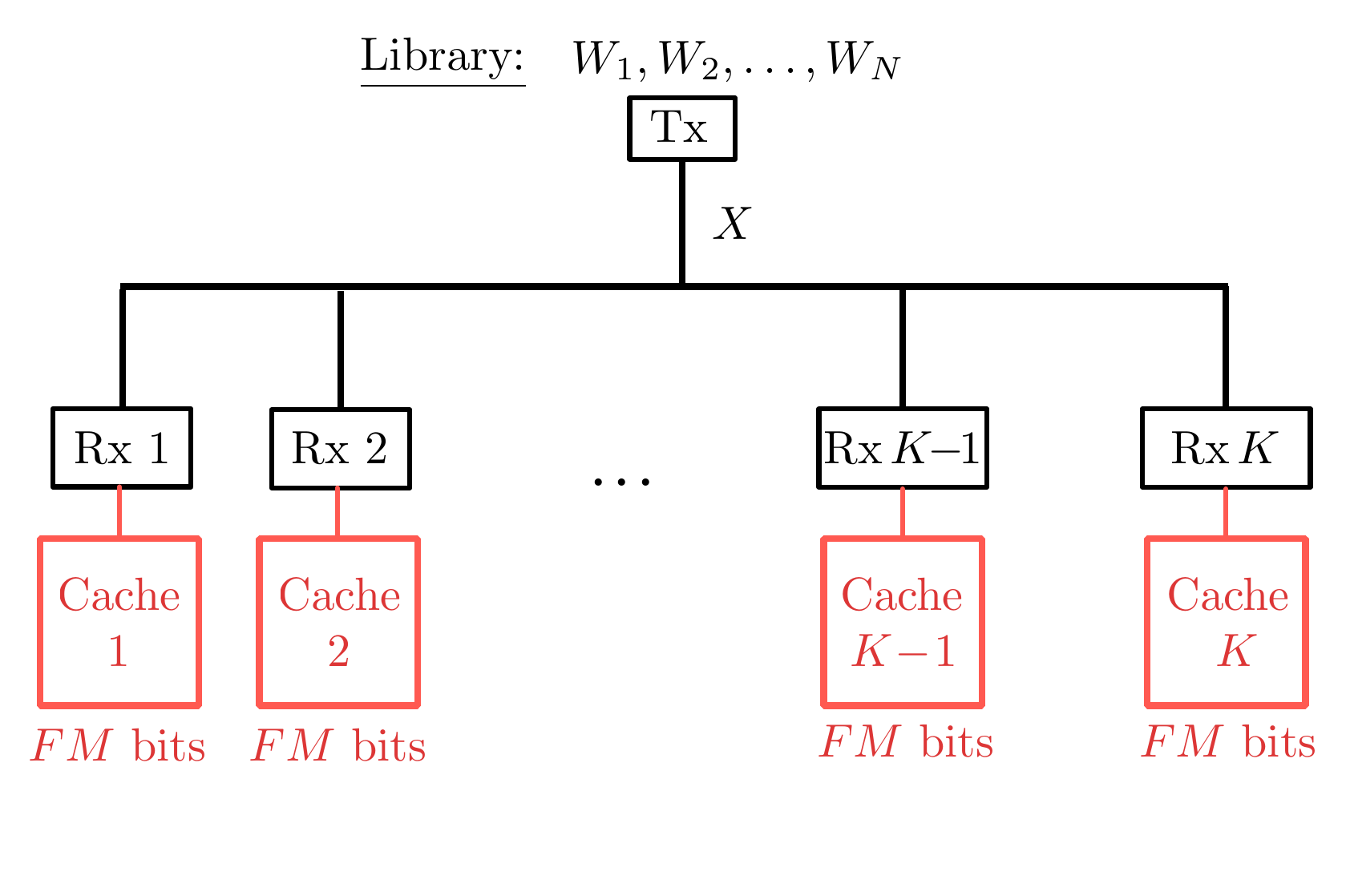}
\caption{Coded caching scenario with $K$ receivers having equal cache size $F M$ bits.}
\label{fig:source_model}
\end{center}
\vspace{-6mm}
\end{figure}

{Consider the communication scenario in  Figure~\ref{fig:source_model}, which includes a single transmitter and $K$ receivers that we term Receivers $1,\ldots, K$.}
The transmitter has a library of $N$ independent messages $W_1,\ldots,W_N$. Each $W_d$ is uniformly distributed over $\{1,\ldots, 2^F \}$ for $F$ a positive integer.
Every receiver is provided with a  cache memory of size $F \M$ bits, {and the  range of interest for $\M$ is 
\begin{equation}
0\leq \M \leq N.
\end{equation}
Here, $\M=0$ means that there is no cache memory in the system  and  $\M= N$ means that every receiver can store all the library in its cache memory.}

Each receiver will demand exactly one message from the library. We denote the demand of Receiver~$k$ by  
\begin{equation}
d_k\in\mathcal{N}:=\{1,\ldots, N\}
\end{equation} and thus the 
message demanded by Receiver~$k$ is $W_{d_k}$. Let
\begin{IEEEeqnarray}{rCl}
\d &:=& (d_1,\ldots, d_K)
\end{IEEEeqnarray}
denote the receivers' demand vector. 

The communications process takes place in two phases, namely the caching phase and the delivery phase.
Caching is done during a period of low network-congestion and before the receivers' demand vector $\d$ is known. More specifically, for  $k\in \{1,\ldots, K\}$,  the transmitter sends  an individual cache message  $\V_k \in\big\{1,\ldots, \big\lfloor 2^{F\M}\big \rfloor\big\}$ to Receiver~$k$. Since $\d$ is unknown at this time, the cache messages will be functions of the entire library. For every $k\in\{1,\ldots, K\}$, we  have\footnote{Alternatively, one could allow the caching functions to depend also on external randomness that does not depend on the library nor the receivers' demands.  {The  rate-memory tradeoff, which is the focus of this paper, is the same under  both assumptions. This can be proved in a similar way as proving that  randomized encoding  does not change the rate-distortion function of memoryless source coding problems.}}
\begin{IEEEeqnarray}{rCl}
\V_k &:=& g_k(W_1, \ldots, W_N),
\end{IEEEeqnarray}
{for some caching function}
\begin{IEEEeqnarray}{rCl}
g_k &:& \{1,\ldots, 2^F\}^N \to \{1,\ldots, \lfloor 2^{F \M}\rfloor \}.\label{eq:caching}
\end{IEEEeqnarray}


 In the delivery phase, the transmitter is given the receivers' demands $\d=(d_1,\ldots, d_K)$, and it generates the delivery-symbol $X$ that is  sent over the common noise-free bit-pipe:
\begin{IEEEeqnarray}{rLl}
X &:= &f_\d(W_1,\ldots,W_N),
\end{IEEEeqnarray}
{for some encoding function}
\begin{IEEEeqnarray}{rCl}
f_\d&\colon& \big\{1,\ldots, 2^F \big\}^N \to  \set{X},\label{eq:encoding}
\end{IEEEeqnarray}
where $\set{X}$ is the delivery alphabet that we will specify shortly. 
We  assume that $\mathbf{d}$ is known to all the receivers (e.g.,~$\d$ can be communicated to the receivers with asymptotically zero transmission rate\footnote{{Alternatively, the desired information could also be sent from the server to the users as part of the subsequent delivery communication, see \cite{Fadlallah}.}}). 

Receiver~$k$, $k\in\{1,\ldots, K\}$, perfectly observes the delivery-symbol $X$, and  thus recovers its desired message as 
\begin{IEEEeqnarray}{rLl}
\hat{W}_k &:=& \varphi_{k,\d}(X, \V_k)
\end{IEEEeqnarray}
{using some decoding function}
\begin{IEEEeqnarray}{rCl}
\varphi_{k,\d} &:& \set{X} \times \{1,\ldots, \lfloor 2^{F \M}\rfloor \} \to \big\{1,\ldots, 2^F\big\}.\label{eq:decoding}
\end{IEEEeqnarray}


We are left with specifying the delivery alphabet $\mathcal{X}$. We distinguish  the {\emph{worst-case} \cite{maddahali_niesen_2014-1} and  \emph{average-case} \cite{niesen_maddahali_2014} scenarios as follows:}
\begin{itemize}
	\item In the \emph{worst-case scenario}, the delivery alphabet $\mathcal{X}$ does not depend on the demand vector $\d$.
	In this scenario, the rate-memory pair $(R,\ \M)$  is  \emph{achievable}
	if for every $\epsilon >  0$ and 
	sufficiently large message lengths~$F$, there {exists a caching function \eqref{eq:caching},  an encoding function \eqref{eq:encoding}, and decoding functions \eqref{eq:decoding}}  for delivery alphabet 
		\begin{equation}\label{eq:X_alphabet}
	\mathcal{X}=\big\{1, \ldots, \lfloor 2^{F(R+\epsilon)}\rfloor\big\},
	\end{equation}
	  so that for each demand vector  $\d\in\set{N}^K$,  every Receiver~$k$, $k\in\{1,\ldots, K\}$, can perfectly reconstruct its desired message:
	\begin{equation}\label{eq:noerror}
	\hat{W}_k = W_{d_k}.
	\end{equation}
	\item In the \emph{average-case scenario}, the delivery alphabet $\set{X}$  depends on the demand vector $\d$. 
			 In this scenario, the rate-memory pair $(R,\ \M)$  is  \emph{achievable}
			if for each demand vector $\d\in\set{N}^K$, any $\epsilon >  0$,
			and sufficiently large message lengths $F$, there {exists a caching function \eqref{eq:caching},  an encoding function \eqref{eq:encoding}, and decoding functions \eqref{eq:decoding}} for 
			 delivery alphabet
		\begin{equation}\label{eq:Xd_alphabet}
		\mathcal{X}_\d=\{1, \ldots, \lfloor 2^{F R_\d}\rfloor\},
		\end{equation}
		so that each Receiver~$k \in \{1,\ldots, K\}$ can perfectly reconstruct its desired message \eqref{eq:noerror} and 
			\begin{IEEEeqnarray}{rCl}
				\frac{1}{N^K} \sum_{\d\in\set{N}^K} R_\d &\leq &R+\epsilon.
			\end{IEEEeqnarray}
\end{itemize}

	The main focus of this paper is on the  \emph{rate-memory tradeoffs} of the worst-case and the average-case scenarios.

\begin{definition}
Given the cache memory size $\M$, we define the \emph{rate-memory tradeoffs} $\RMworst$ and $\RMavg$ as the infimum of all rates $R$ such that the rate-memory pair $(R,\ \M)$ is achievable for the worst-case and  average-case scenarios, respectively. 
\end{definition}


\section{Main Results}

Define $\bar{N} := \min\{K,N\}$ and $\bar{\mathcal{N}}:=\{1,2,\ldots,\bar{N}\}$.

\subsection{Worst-Case Scenario}
Our first result is a lower bound on the rate-memory tradeoff in the worst-case scenario and it is proved in Section~\ref{subsec:worst}. The bound can also be extracted from the converse result for general degraded broadcast channels  in \cite{SaeediWiggerYener-2016}.
\begin{theorem}\label{thm:conv_worst}
For all $\M\in[0,N)$, 
\begin{equation}
\RMworst \geq \Rlowerworst(\M),
\end{equation}
where 
\begin{IEEEeqnarray}{rCl}
\Rlowerworst(\M) &:=& \max\Bigg\{\max_{\ell \in \bar{\mathcal{N}}} \left[
\ell - \M \frac{\ell^2}{N} \right] , \nonumber \\
&& \hspace{1.05cm}\max_{\ell \in \bar{\mathcal{N}}} \left[ \ell - \M \sum_{j=1}^\ell \frac{j}{N-j+1} \right] \Bigg\}.\label{eq:low1} \IEEEeqnarraynumspace
\end{IEEEeqnarray}
\end{theorem}

\begin{figure}[t!]

\includegraphics[scale=0.3]{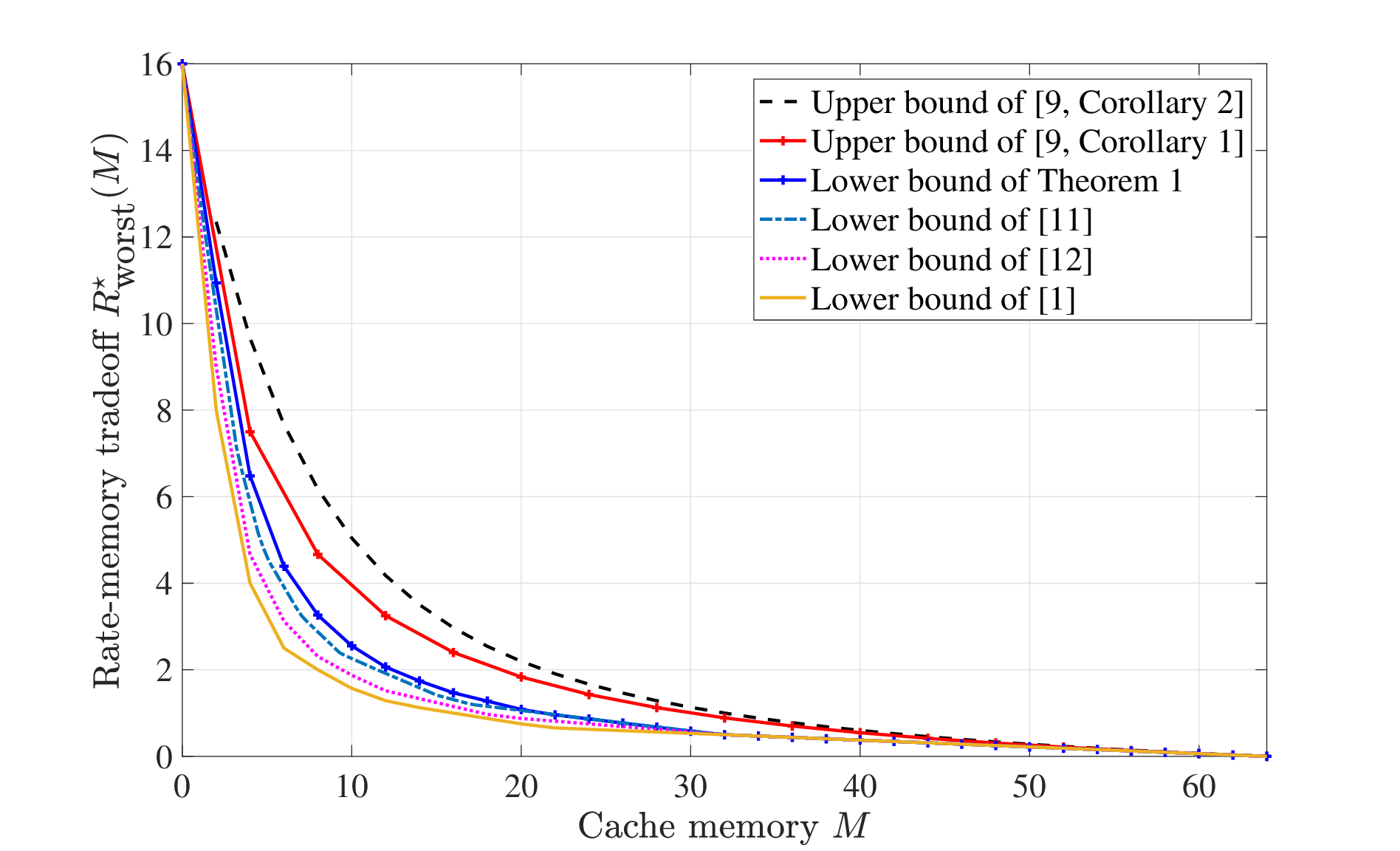}
\caption{Upper and lower bounds on $R_{\textnormal{worst}}^\star(\M)$ for $K=16$ and $N=64$.}
\label{fig:source_worst}
\vspace{-0.1cm}
\end{figure}

Figure~\ref{fig:source_worst} compares our new lower bound on $\RMworst$ with  the existing lower bounds in \cite{maddahali_niesen_2014-1,ghasemi_ramamoorthy,sengupta_tandon_clancy-2015}. 
The figure also shows  upper bounds from \cite{YuMA:16}.  The solid red upper bound is for centralized caching, as considered in this paper. The dashed black upper bound is for decentralized caching. For simplicity, the latter upper bound is used to derive gap results as stated in Theorem~\ref{thm:gap_worst} below. 

\begin{theorem} \label{thm:gap_worst}
Irrespective of the number of users $K$, the library size $N$, and the  memory size $\M\in[0,N)$:
\begin{IEEEeqnarray}{rCl}
\frac{R^\star_{\sf worst}(\M)}{\Rlowerworst(\M)} &\le& \max_{\ell \in \mathbb{Z}^+} \max_{a\in(0,1)} \phi(a,\ell), 
\end{IEEEeqnarray}
where 
\begin{IEEEeqnarray}{rCl}
\phi(a,\ell) &:=& \frac{\frac{a(\ell+1)}{(1-a)\ell}\left(1-\left(\frac{\ell+a}{\ell+1}\right)^{\ell/a}\right)}{1 - (1-a)\sum_{j=0}^{\ell-1} \frac{1}{\ell-a j}}.
\end{IEEEeqnarray}
\end{theorem}
\begin{IEEEproof}
See Section~\ref{sec:gap_worst}.
\end{IEEEproof}

\begin{figure}[t!]
\centering
\includegraphics[width=0.51\textwidth]{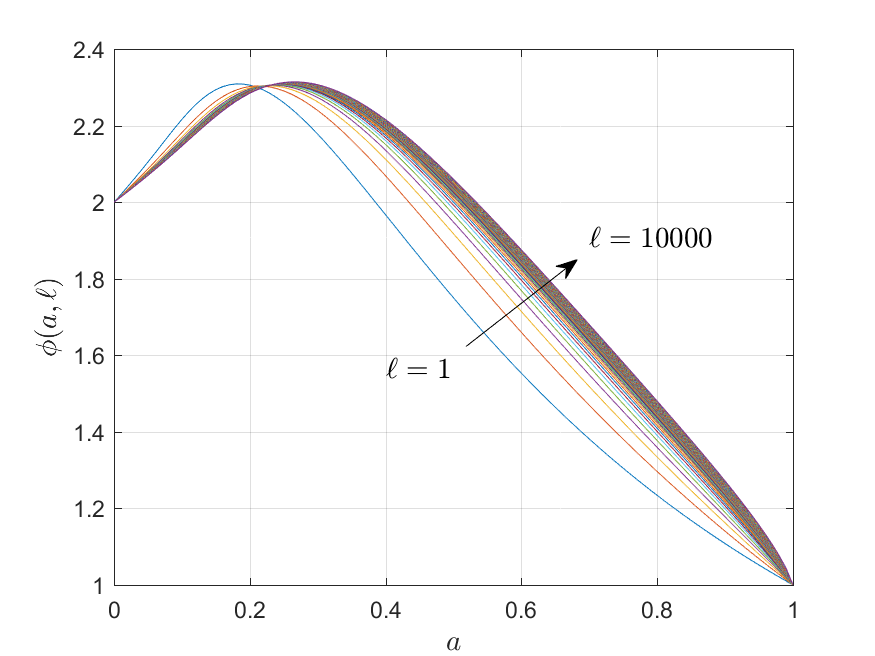}
\caption{The functions $\phi(a, \ell)$ for $a\in(0,1)$ and $\ell=1,\dots, 10^4$.}
\label{fig:plot_phi}
\end{figure}

\begin{figure}[t!]
\centering
\includegraphics[width=0.5\textwidth]{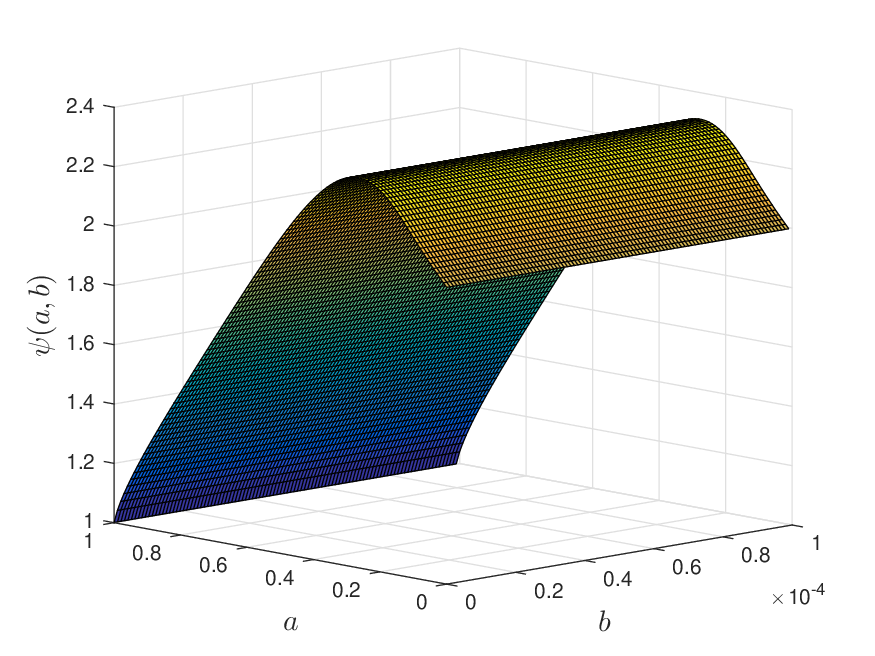}
\caption{The function $\psi(a, b)$ over $(a,b)\in(0,1)\times (0,10^{-4})$.}
\label{fig:plot_psi}
\end{figure}

\begin{remark}\label{rem:2315}
For any $\ell \in \mathbb{Z}^+$, the function $a\mapsto \phi(a,\ell)$ is continuous and bounded over $(0,1)$, see also Figure~\ref{fig:plot_phi}. Numerical evaluations\footnote{All numerical evaluations in this paper are performed by applying the MATLAB function \textsf{fmincon} with the sequential quadratic programming (SQP) method.} show that for $\ell \in\{1,\ldots,10^4\}$:
\begin{equation}
\max_{a\in(0,1)} \phi(a,\ell) \leq 2.315.
\end{equation}
Moreover, 
\begin{equation}\label{eq:phipsi}
\max_{\ell\in\mathbb{Z}^+\colon \ell > 10^4}\; \max_{a\in(0,1)} \phi(a,\ell) \leq  \max_{b\in(0,10^{-4})}\; \max_{a\in(0,1)} \psi(a,b),
\end{equation}
where
\begin{equation}\label{eq:psi}\psi(a,b):= \frac{\frac{a(1+b)}{1-a}\left(1-\left(\frac{1+ab}{1+b}\right)^{\frac{1}{ab}}\right)}{1 - \frac{(1-a)b}{1-a+ab} + \frac{1-a}{a}\ln\left(1-a+ab\right)}.
\end{equation}
The function $(a,b)\mapsto\psi(a,b)$ is continuous and bounded over $(0,1)\times(0,10^{-4})$, see also Figure~\ref{fig:plot_psi}. Numerical evaluations show that 
\begin{equation}
 \max_{b\in(0,10^{-4})}\; \max_{a\in(0,1)} \psi(a,b) \leq 2.315.
 \end{equation}
\end{remark}
\begin{IEEEproof}
Inequality~\eqref{eq:phipsi} is proved in Section~\ref{sec:proofremark}.
\end{IEEEproof}

\subsection{Average-Case Scenario}
\begin{theorem}\label{thm:conv_avg}
For all $\M\in[0,N)$, 
\begin{IEEEeqnarray}{rCl}
\RMavg &\geq& \Rloweravg(\M), 
\end{IEEEeqnarray}
where 
\begin{IEEEeqnarray}{ll}
&\Rloweravg(\M) \nonumber \\
&:= \max \Bigg\{ \max_{\ell\in\{1,\ldots,K\}} \left[\Big( 1 - \Big(1- \frac{1}{N}\Big)^\ell \Big) (N -\ell \M) \right] ,\nonumber \\
 & \hspace{1.5cm}\max_{\ell\in\{1,\ldots,K\}}  \left[\Big( 1 - \Big(1- \frac{1}{N}\Big)^\ell \Big)N - \frac{\ell(\ell+1)}{2N}\M \right] \Bigg\}. \nonumber\\
\end{IEEEeqnarray}
\end{theorem}
\begin{IEEEproof}
See Section~\ref{subsec:avg}.
\end{IEEEproof}

\begin{figure}[t!]
\centering
\includegraphics[scale=0.3]{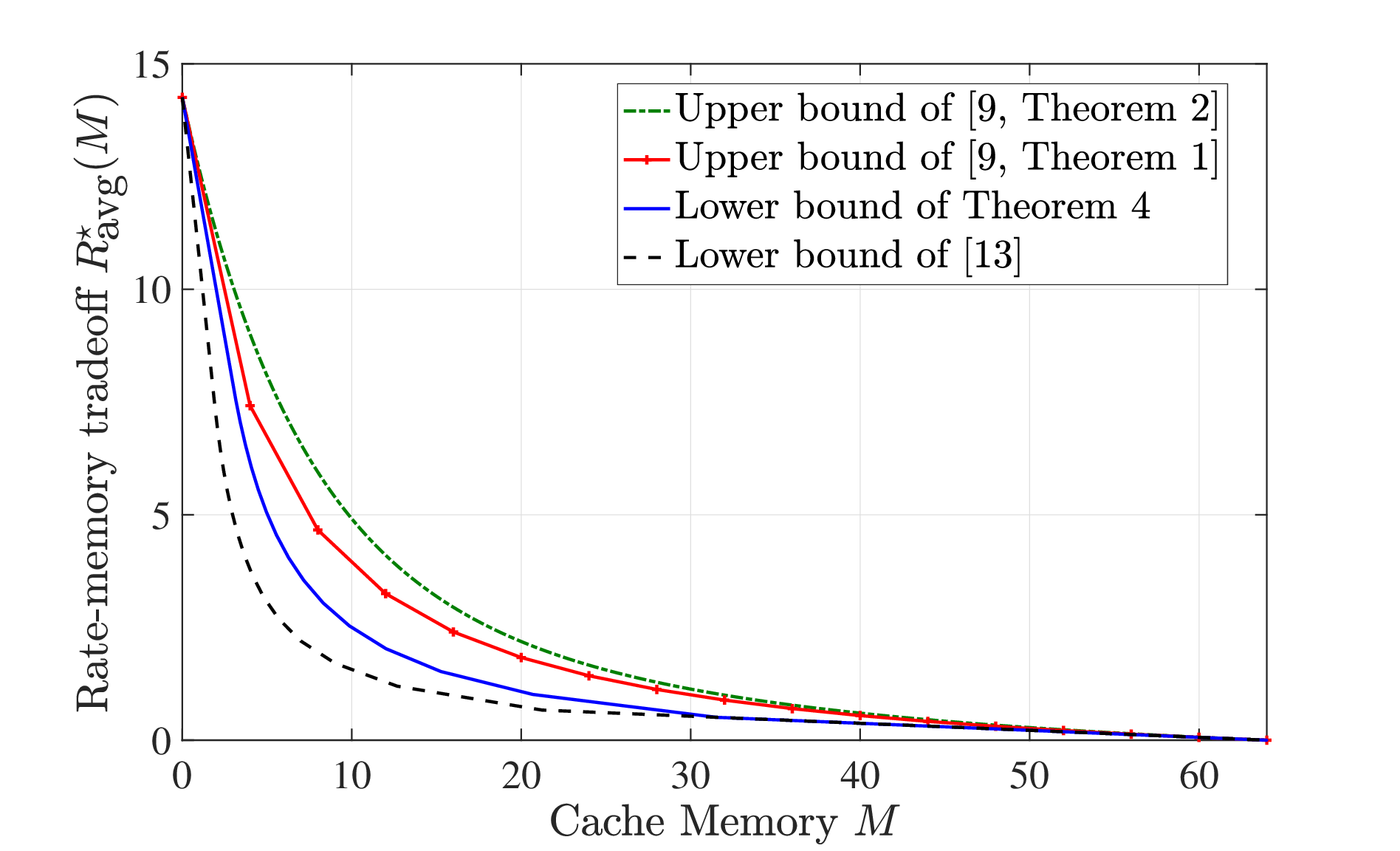}
\caption{Upper and lower bounds on $R_{\textnormal{avg}}^\star(\M)$ for $K=16$ and $N=64$.}
\label{fig:source_avg}
\end{figure}

Figure~\ref{fig:source_avg} compares this new lower bound on $\RMavg$ with  the existing lower bounds in \cite{wanglimgastpar-2016} and the upper bounds in \cite{YuMA:16}.  The  solid red upper bound is for centralized caching, as considered in this paper. The  dotted green upper bound is for decentralized caching and also from \cite{YuMA:16}.

As the following theorem and remark show, the multiplicative gap between the lower bound of Theorem~\ref{thm:conv_avg} and $\RMavg$ is at most~2.507.

\begin{theorem} \label{thm:gap_avg}
Irrespective of the number of users $K$, the library size $N$, and the  memory size $\M\in[0,N)$:
\begin{IEEEeqnarray}{rCl}
\frac{\RMavg}{\Rloweravg(\M)} &\le& \max_{u\in(0,1]}\max_{v\in(0,1/2]} \eta(u,v), 
\end{IEEEeqnarray}
where 
\begin{IEEEeqnarray}{ll}
& \lefteqn{\eta(u,v)} \nonumber \\
&:= \frac{\left(u+v-v\left(1-v\right)^{\frac{u}{v}}\right)\left(1-\left(1-\frac{v}{u+v}\left(1-v\right)^{\frac{u}{v}}\right)^{\frac{1}{v}}\right)}{\left(1-v\right)^{\frac{u}{v}}\left(1-\left(1+\frac{u}{2}\right)\left(1-v\right)^{\frac{u}{v}}\right)}.\nonumber \\
\end{IEEEeqnarray}
\end{theorem}
\begin{IEEEproof}
See Section~\ref{sec:gap_avg}.
\end{IEEEproof}

\begin{remark}\label{rem:2507}
	The function $\eta(u,v)$ is continuous and bounded over $(0,1]\times (0, 1/2]$. 
Numerical evaluations show that  
\begin{equation}
 \max_{u\in(0,1]}\max_{v\in(0,1/2]} \eta(u,v) \leq 2.507.
\end{equation}
\end{remark}

\begin{figure}[t!]
\centering
\includegraphics[width=0.5\textwidth]{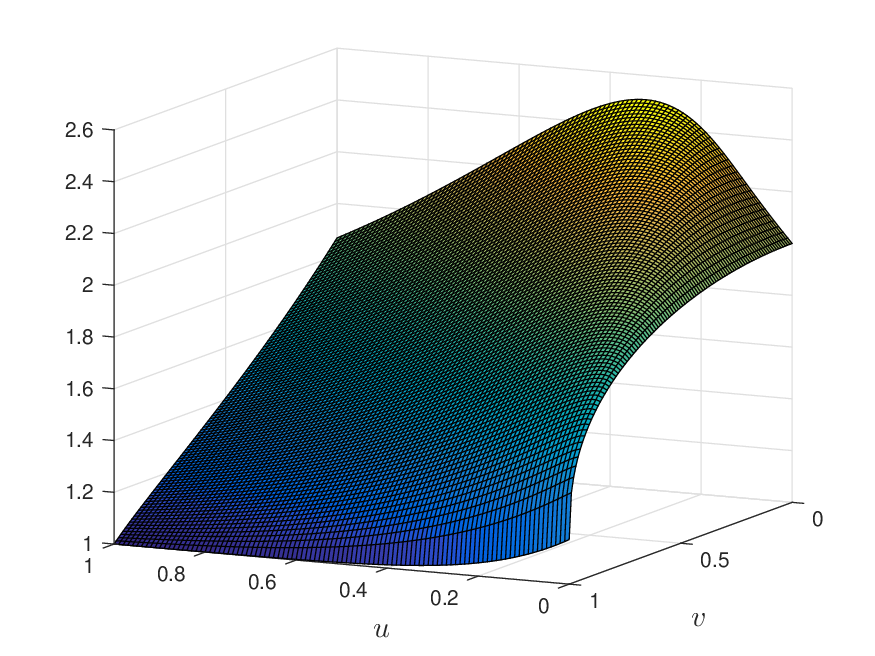}
\caption{The function $\eta(u,v)$ over $(u,v)\in(0,1)\times (0,1)$.}
\label{fig:plot_eta}
\end{figure}

\section{Proof of Theorems~\ref{thm:conv_worst} and \ref{thm:conv_avg}}\label{sec:conv}

\subsection{Auxiliary Lemmas}
The following two lemmas will be used in the proofs of Theorems~\ref{thm:conv_worst} and \ref{thm:conv_avg}. 

Lemma \ref{lem:fixedd} (see below) is stated for the average-case scenario. It also  readily applies to the worst-case scenario if rate $R_\d$ is replaced by $\R$.
\begin{lemma}\label{lem:fixedd}
	Fix a number $\ell\in\bar{\mathcal{N}}$ and a demand vector $\d\in\set{N}^K$ whose first $\ell$ entries are $d_1,\ldots, d_\ell$. Fix also a small $\epsilon>0$ and assume a sufficiently large $F$ with  caching, encoding, and decoding functions so that \eqref{eq:noerror} holds for all $k\in\{1,\ldots, K\}$. Then, 
\begin{equation}\label{eq:fixedd}
R_\d+\epsilon \geq \kappa_{\d}(\ell)-  \frac{1}{F} \sum_{k=1}^\ell  I(W_{d_k};\V_1, \ldots, \V_k| W_{d_1}, \ldots, W_{d_{k-1}}),
\end{equation}
where $\kappa_\d(\ell)$ denotes the number of distinct demands for Receivers~$1,\ldots, \ell$:
\begin{equation}
\kappa_{\d}(\ell):=| \{d_1,\ldots, d_\ell \}|.
\end{equation}
\end{lemma} 
\begin{IEEEproof} 
For any $k\in\{1,\ldots, \ell\}$:
\begin{IEEEeqnarray}{rCl}\label{eq:fix_k}
\lefteqn{I(X;W_{d_k}| \V_1, \ldots, \V_k, W_{d_1}, \ldots, W_{d_{k-1}})}\qquad  \nonumber \\
  & \stackrel{(a)}{=} & H(W_{d_k}| \V_1, \ldots, \V_k, W_{d_1}, \ldots, W_{d_{k-1}})  \nonumber \\
  & = & H(W_{d_k}|W_{d_1}, \ldots, W_{d_{k-1}}) \nonumber \\
  &&  -I(W_{d_k};\V_1, \ldots, \V_k| W_{d_1}, \ldots, W_{d_{k-1}})  \nonumber \\
  &\stackrel{(b)}{=} & F \cdot \mathbbm{1}\big\{d_k \notin \{d_1,\ldots, d_{k-1}\}\big\} \nonumber \\
   &&  -I(W_{d_k};\V_1, \ldots, \V_k| W_{d_1}, \ldots, W_{d_{k-1}}),\IEEEeqnarraynumspace
\end{IEEEeqnarray}
{where $ \mathbbm{1}\big\{d_k \notin \{d_1,\ldots, d_{k-1}\}$ denotes the indicator function that is $1$ if $d_k$ is not in $\{d_1,\ldots, d_{k-1}\}$ and is $0$ otherwise. Moreover,  (a) holds because \eqref{eq:noerror} implies that $H(W_{d_k}| X, \V_1, \ldots, \V_k, W_{d_1},\ldots, W_{d_{k-1}})=0$ and  (b)} holds by the independence of the messages and because $H(W_d)=F$ for any $d\in\set{N}$.

On the other hand, 
\begin{IEEEeqnarray}{rCl}\label{eq:HX}
\lefteqn{\sum_{k=1}^\ell I(X;W_{d_k}| \V_1, \ldots, \V_k, W_{d_1}, \ldots, W_{d_{k-1}})}\qquad  \nonumber \\
&\leq & \sum_{k=1}^\ell I(X;W_{d_k}, \V_k| \V_1, \ldots, \V_{k-1}, W_{d_1}, \ldots, W_{d_{k-1}}) \nonumber \\
& = & I(X; W_{d_1},\ldots, W_{d_{\ell}}, \V_1,\ldots, \V_\ell)  \nonumber \\
&\leq & H(X)  \nonumber \\
& \stackrel{(a)}{\leq}  & F (R_\d+\epsilon),
\end{IEEEeqnarray}
{where $(a)$ holds by \eqref{eq:X_alphabet}.}
Combining \eqref{eq:fix_k} and \eqref{eq:HX} establishes the lemma.
\end{IEEEproof}

\begin{lemma} \label{prop:han}
Let $L$ be a positive integer, $(A_1,\ldots, A_L)$ be a random independent $L$-tuple, and $\V$ be a random variable arbitrarily correlated with $A_1,\ldots, A_L$. For any subset $\set{S}\subseteq \{1,\ldots, L\}$, denote  the subset $\{A_s,\ s\in\mathcal{S}\}$ by  $A_{\mathcal{S}}$.
Then, for all $l\in\{1,\ldots, L\}$, 
\begin{align}
{1\over {L \choose l }}\sum_{\substack{\mathcal{S}\subseteq\{1,\ldots,L\}:\\|\mathcal{S}|=l}}I(A_{\mathcal{S}};\V)\leq \frac{l}{L}I(A_1,\ldots,A_L;\V).\label{Han}
\end{align}
\end{lemma}
\begin{IEEEproof}
Consider any $l\in\{1,\ldots, L\}$. We have 
\begin{IEEEeqnarray}{ll}
& \lefteqn{{1\over {L \choose l }}\sum_{\substack{\mathcal{S}\subseteq\{1,\ldots,L\}:\\|\mathcal{S}|=l}}I(A_{\mathcal{S}};\V)} \nonumber \\
&\overset{(a)}{=} {1\over {L \choose l }}\sum_{\substack{\mathcal{S}\subseteq\{1,\ldots,L\}:\\|\mathcal{S}|=l}}  \;\; \sum_{j\in\mathcal{S}} H(A_j) \nonumber \\&\quad \;  - {1\over {L \choose l }}\sum_{\substack{\mathcal{S}\subseteq\{1,\ldots,L\}:\\|\mathcal{S}|=l}} H(A_{\mathcal{S}}|\V) \nonumber \\
&= {{L-1 \choose l-1 }\over {L \choose l }} \sum_{j=1}^L H(A_j) - {1\over {L \choose l }}\sum_{\substack{\mathcal{S}\subseteq\{1,\ldots,L\}:\\|\mathcal{S}|=l}} H(A_{\mathcal{S}}|\V) \nonumber \\
&\overset{(b)}{\le} \frac{l}{L} \sum_{j=1}^L H(A_j) - \frac{l}{L} H(A_1,\ldots,A_L|\V) \nonumber\\
&\overset{(c)}{=} \frac{l}{L} I(A_1,\ldots,A_L;\V), 
\end{IEEEeqnarray}
where $(a)$ and $(c)$ follow since $A_1,\ldots, A_L$ are independent and $(b)$ follows from the generalized Han Inequality (see~\cite[Theorem 17.6.1]{CoverThomas}).
\end{IEEEproof}

\subsection{Proof of Theorem~\ref{thm:conv_worst}} \label{subsec:worst}

Fix $\ell\in\bar{\mathcal{N}}$ and restrict attention to Receivers~$1,\ldots, \ell$ and their cache memories. Let $\mathcal{Q}_\ell^{\textnormal{dist}}$ be the set of all \emph{ordered} $\ell$-dimensional demand vectors $(d_1,\ldots, d_\ell)$ with all distinct entries. So, 
\begin{equation}
|\mathcal{Q}_\ell^{\textnormal{dist}}|= {N \choose \ell}\ \ell! .
\end{equation}

Notice that for $\d\in\Qell$, we have $\kappa_\d(\ell)=\ell$, and averaging Inequality~\eqref{eq:fixedd}\footnote{In \eqref{eq:fixedd} $R_\d$ needs to be replaced by $R$ because here we consider a worst-case scenario.} over all demand vectors $\d\in\Qell$  yields the following inequality:
\begin{equation}\label{eq:rate_worst}
R +\epsilon \geq \ell -  \sum_{k=1}^\ell \alpha_k,
\end{equation}
 where
\begin{subequations}\label{defgen}
\begin{IEEEeqnarray}{rCl}
{\alpha}_{1}&:= \frac{1}{{N \choose \ell}{{\ell!}}}\sum_{\d \in \Qell }\frac{1}{F} I(W_{d_1};\V_{1}), 
\end{IEEEeqnarray}
and for $k=2,\ldots, \ell$:
\begin{IEEEeqnarray}{rCl}
{\alpha}_{k}&:= \frac{1}{{N \choose \ell}{{\ell!}}}\sum_{\d \in \Qell}\frac{1}{F} I(W_{d_k};\V_{1}, \ldots, \V_{k}|W_{d_1}, \ldots, W_{d_{k-1}}).
\end{IEEEeqnarray}
\end{subequations}

We now upper bound the terms $\sum_{k=1}^\ell \alpha_k$ that appear on the right hand side of \eqref{eq:rate_worst}. In particular, we prove the following lemma in Appendix~\ref{app:lem3}.

\begin{lemma}\label{lem:alphas}
Parameters  $\alpha_1, \ldots, \alpha_\ell$ satisfy 
\begin{align}
\label{eq:cons3prime}
\sum_{j=1}^\ell {\alpha}_{k} &  \leq  \min\bigg\{ \frac{\ell^2}{N}\M , \ \sum_{j=1}^\ell \frac{j \M}{N-j+1}\bigg\}.
\end{align} 
\end{lemma}
\begin{IEEEproof}{See Appendix~\ref{app:lem3}.}
	\end{IEEEproof}
	
	Inserting \eqref{eq:cons3prime} into \eqref{eq:rate_worst}, we obtain
			\begin{equation}
R+\epsilon \geq \ell -  \leq  \min\bigg\{ \frac{\ell^2}{N}\M , \ \sum_{j=1}^\ell \frac{j \M}{N-j+1}\bigg\}.  
		\end{equation} 
		Finally, letting
		$\epsilon\to0$ concludes the proof.

\subsection{Proof of Theorem~\ref{thm:conv_avg}} \label{subsec:avg}

For any $\ell\in\{1,\ldots, K\}$, let  $\Qrepell$ be the set of all ordered length-$\ell$ vectors $(d_1,\ldots, d_\ell)\in\set{N}^\ell$, where repetitions are allowed. Notice that:
\begin{equation}
|\Qrepell|= N^\ell.
\end{equation}
Recall also that in the average-case scenario under investigation, the demand vector $\d:=(d_1, \ldots, d_K)$ is uniform over~$\Qrep_K$. Let $\D := (D_1,\ldots,D_K)\sim$ Uniform($\mathcal{N}^K$). 

Fix now an $\ell\in\{1,\ldots, K\}$, and average Inequality~\eqref{eq:fixedd} over all demand vectors $\d\in\Qrep_K$. This yields:
\begin{equation}\label{eq:rate_avg}
\E_{\D}[ R_\D+ \epsilon] \geq \E_{\D}[ \kappa_\D(\ell)] -  \sum_{k=1}^\ell \beta_{k},
\end{equation}
where
\begin{subequations}\label{defgen_avg}
\begin{IEEEeqnarray}{rCl}
{\beta}_{1}&:=  \frac{1}{F} I(W_{D_1};\V_{1}|\D) ,
\end{IEEEeqnarray}
and for $k=2,\ldots, \ell$:
\begin{IEEEeqnarray}{rCl}
{\beta}_{k}&:=
\frac{1}{F} I(W_{D_k};\V_{1}, \ldots, \V_{k}|W_{D_1}, \ldots, W_{D_{k-1}},\D).\IEEEeqnarraynumspace
\end{IEEEeqnarray}
\end{subequations}

{We can bound the terms in \eqref{eq:rate_avg} with the following two lemmas.}

\begin{lemma}
\begin{equation}
 \E_\D\big[\kappa_\D(\ell)\big] =N \Big(1-\Big(1-\frac{1}{N}\Big)^\ell\Big).
 \end{equation}
\end{lemma}
\begin{IEEEproof}
	{
\begin{IEEEeqnarray}{rCl}\lefteqn{
 \E_\D\big[\kappa_\D(\ell)\big] } \nonumber \\
 & = & \E_\D\Bigg[ \sum_{k=1}^\ell \mathbbm{1} \big\{D_k \notin \{D_1,\ldots, D_{k-1}\}\big\}\Bigg] \nonumber \\
  &=& \sum_{k=1}^\ell\E_\D\Big[  \mathbbm{1} \big\{D_k \notin \{D_1,\ldots, D_{k-1}\}\big\}\Big] \nonumber \\
    & \  \stackrel{(a)}{=} & \sum_{k=1}^\ell \sum_{j=1}^{N} \frac{1}{N}\E_\D\Big[  \mathbbm{1} \big\{j \notin \{D_1,\ldots, D_{k-1}\}\big\} \Big|D_k =j \Big] \nonumber \\
    & \stackrel{(b)}{=} & \sum_{k=1}^\ell \Big(1- \frac{1}{N}\Big)^{k-1} \nonumber \\
    &\stackrel{(c)}{=}& N\left(1 -  \Big(1- \frac{1}{N}\Big)^{\ell}\right),
\end{IEEEeqnarray}
where $(a)$ holds by the  law of total expectation and because $D_k$ is uniform over $\{1,\ldots, N\}$; $(b)$ holds because $D_1,\ldots, D_{k-1}$ are i.i.d. and uniform over $\{1,\ldots, N\}$; and $(c)$ follows  by the formula of a geometric sum.}
\end{IEEEproof}

\begin{lemma}\label{lem:betas}
Parameters  $\beta_1, \ldots, \beta_\ell$ satisfy 
\begin{align}
\sum_{j=1}^\ell {\beta}_{j} &  \leq  \min\bigg\{ \E_\D\big[\kappa_\D(\ell)\big]\cdot \frac{\ell \M}{N}, \ \sum_{j=1}^\ell \frac{j \M}{N}\bigg\}.
\end{align} 
\end{lemma}
\begin{IEEEproof}{See Appendix~\ref{app:lem11}.}
	\end{IEEEproof}
	
{
Combining the above two lemmas with \eqref{eq:rate_avg}, we obtain:
\begin{IEEEeqnarray}{rCl}\label{eq:rate_avg2}
\E_{\D}[ R_\D+ \epsilon]& \geq& N\left(1 -  \Big(1- \frac{1}{N}\Big)^{\ell}\right) \nonumber \\
 & & -  \min\bigg\{ \E_\D\big[\kappa_\D(\ell)\big]\cdot \frac{\ell \M}{N}, \ \sum_{j=1}^\ell \frac{j\M}{N}\bigg\}.
\end{IEEEeqnarray} 
Finally, letting 
 $\epsilon\to 0$ concludes the proof.}

\section{Proof of the Gap-Results in Theorem~\ref{thm:gap_worst} and Remark~\ref{rem:2315}}\label{sec:gap_worst}
\subsection{Proof of Theorem~\ref{thm:gap_worst}}
We wish to uniformly bound the gap 
\begin{equation}
\xi(K,N,\M) := \frac{\RMworst}{\Rlowerworst(\M)},
\end{equation}
irrespective of $K,N\geq 1$ and $\M\in[0,N)$.

Recall the achievable rate-memory tradeoff from \cite[Corollary 2]{YuMA:16} and denote it by $R_{\sf YMA}(K, N, \M) $. For any pair of positive integers $K, N \geq 1$, we have 
\begin{IEEEeqnarray}{ll}
& R_{\sf YMA}(K, N, \M) \nonumber \\
&:= 
\begin{cases}
\bar{N} & \text{ if } \M=0, \\
\frac{N-\M}{\M}\left(1-\left(1-\frac{\M}{N}\right)^{\bar{N}}\right) & \text{ if } \M \in(0, N).
\end{cases}
\end{IEEEeqnarray}
Since $ R_{\sf YMA}(K, N, \M)$ upper bounds the rate-memory tradeoff under a \emph{decentralized caching} assumption \cite{YuMA:16}, it must also upper bound the rate-memory tradeoff under \emph{centralized caching} as considered here. (In fact, decentralized caching imposes additional constraints on the caching functions $g_{k}$ compared to our setup here.)  Thus, for any number of users $K$ and files $N$:
\begin{IEEEeqnarray}{rCl} 
\label{eq:YMAdec}
\RMworst &\le&R_{\sf YMA}(K, N, \M), \quad \M\in[0,N).
\end{IEEEeqnarray}
We thus have
\begin{IEEEeqnarray}{rCl} \label{eq:new_gap}
\xi(K,N,\M) &\leq& \frac{ R_{\sf YMA}(K, N, \M)}{\Rlowerworst(\M)}  \nonumber \\
&\leq & \frac{ R_{\sf YMA}(K, N, \M)}{\RlowerworstUnd(K,N,\M)},  \IEEEeqnarraynumspace
\end{IEEEeqnarray}
where we define
\begin{equation}\label{eq:deno}
\RlowerworstUnd(K,N,\M):=\max_{\ell\in\bar{\mathcal{N}}} \sum_{j=1}^{\ell} \left(1-\frac{j\M}{N-j+1}\right)\end{equation}
and the second inequality holds because for all $K,N,\M$: 
\begin{IEEEeqnarray}{rCl}
\Rlowerworst(\M)& \geq & \max_{\ell\in\bar{\mathcal{N}}} \left[ \ell - \sum_{j=1}^{\ell} \frac{j\M}{N-j+1} \right] \nonumber \\
& = & \RlowerworstUnd(K,N,\M). 
\end{IEEEeqnarray}

{We have a closer look at the function $\RlowerworstUnd(K,N,\M)$. Define 
	\begin{IEEEeqnarray*}{rCl}
		\M_i &:=& \begin{cases}
			\frac{N-i}{i+1} & \text{ if } i\in\{0,1,\ldots,\bar{N}-1\}, \\
			0 & \text{ if } i = \bar{N}.
		\end{cases}
	\end{IEEEeqnarray*} 
	and notice that
	\begin{equation}0=\M_{\bar{N}} < \M_{\bar{N}-1} < \cdots < \M_0 = N.
	\end{equation}
	In Appendix~\ref{app:piecewise} it is shown that for each $i \in \bar{N}$:
	\begin{equation}\label{eq:equiv1}
\RlowerworstUnd(K,N,\M) = i - M \cdot \sum_{j=1}^{i} \frac{
j}{N-j+1}, \quad M\in[M_{i},M_{i-1}].\end{equation}
So,  for given $K,N$, the function $\RlowerworstUnd(K,N,\M)$   is piecewise-linear with $\bar{N}$ line segments over the intervals
\begin{subequations}\label{eq:intervals}
	\begin{IEEEeqnarray}{rCl}
		\ [\M_{i}, \M_{i-1}], & &\qquad i \in\{2, \ldots, \bar{N}\},  \label{eq:intervals_a}\\
		\ [\M_1, \M_0).&&
	\end{IEEEeqnarray}
\end{subequations}
}

We next upper bound $R_{\sf YMA}(K,N,\M)$ by a function $\overline{R}_{\sf YMA}(K,N,\M)$ that is piecewise-linear over the same intervals  \eqref{eq:intervals}. 
Specifically, for every $\ell\in\bar{\mathcal{N}}$, define for $\M \in [\M_\ell, \M_{\ell-1})$:
{\begin{IEEEeqnarray}{rCl}
\overline{R}_{\sf YMA}(K,N,\M) &:=&  \frac{  \M_{\ell-1}-\M}{\M_{\ell-1}-\M_{\ell}} R_{\sf YMA}(K,N,\M_{\ell})\nonumber \\
 &&  +   \frac{ \M- \M_\ell}{\M_{\ell-1}-\M_\ell}  R_{\sf YMA}(K,N,\M_{\ell-1}). \IEEEeqnarraynumspace
\end{IEEEeqnarray}}
Notice  that 
\begin{equation}\overline{R}_{\sf YMA}(K,N,\M_\ell)
= R_{\sf YMA}(K,N,\M_\ell), \quad \forall \ell\in\{1, \ldots, \bar{N}\}, \end{equation}
whereas for general $\M\in[0,N)$:
\begin{equation}\label{eq:new_upper}
\overline{R}_{\sf YMA}(K,N,\M) \geq {R}_{\sf YMA}(K,N,\M),
\end{equation}
because {for fixed values of $K,N$ the function $R_{\sf YMA}(K,N,\M)$ is convex in $\M$.}

Plugging \eqref{eq:new_upper} into \eqref{eq:new_gap}, we obtain: 
\begin{equation}\label{eq:newest_gap}
\xi(K,N,\M) \leq \frac{\overline{R}_{\sf YMA}(K,N,\M) }{ \RlowerworstUnd(K,N,\M)} =: \Xi(K,N,\M).
\end{equation} 
{Now, since the upper bound $\Xi(K,N,\M)$  is continuous and bounded in $\M\in[0,N)$ and because it is quasiconvex\footnote{A linear-fractional function is always quasiconvex  \cite{Boyd:04}.} in $\M$,  the maximum of $\Xi(K,N,\M)$ over each of the $\bar{N}-1$ \emph{closed} intervals in \eqref{eq:intervals_a} is attained at one of the two  boundary points of the interval. That means, 
\begin{IEEEeqnarray}{rCl}\label{eq:finite_set1}
	\max_{\M\in[0,\M_1]} \xi(K,N,\M) & \leq &  \max_{\ell\in\bar{\mathcal{N}}}  \Xi(K,N,\M_\ell)
\end{IEEEeqnarray}
Consider now the half-open interval $\M\in[\M_1, \M_0)$.} Since $\overline{R}_{\sf YMA}(K,N,\M_{0})= \RlowerworstUnd(K,N,\M_0)=0$,  for $\M\in[\M_1, \M_0)$:
\begin{IEEEeqnarray}{rCl}
\lefteqn{\Xi(K,N,\M) } \quad \nonumber \\
& = &\frac{ {\frac{\M_0-\M}{\M_0- \M_1}} \overline{R}_{\sf YMA}(K,N,\M_{1})+ {\frac{\M- \M_1}{\M_0- \M_1} }\cdot 0}{ {  \frac{\M_0-\M}{\M_0- \M_1} } \RlowerworstUnd(K,N,\M_{1})+   { \frac{\M- \M_1}{ \M_0-\M_1}}\cdot 0} \nonumber \\
&=& \frac{\overline{R}_{\sf YMA}(K,N,\M_{1})}{\RlowerworstUnd(K,N,\M_1) }\nonumber \\
&=& \Xi(K,N,\M_1),
\end{IEEEeqnarray}
{and $\Xi(K, N, \M)$ is constant over  $[\M_1, \M_0)$. So, trivially the maximum is achieved for $\M=\M_1$. Combined with \eqref{eq:finite_set1},  this yields:
\begin{IEEEeqnarray}{rCl}\label{eq:finite_set}
\max_{\M\in[0,N)} \xi(K,N,\M) & \leq & \max_{\ell\in\bar{\mathcal{N}}}  \Xi(K,N,\M_\ell). 
\end{IEEEeqnarray}}

{We continue to bound the right-hand side of \eqref{eq:finite_set}.} Irrespective of $K,N \in \mathbb{Z}^+$, we have:
\begin{IEEEeqnarray}{rCl}\label{eq:ratio1}
\Xi(K,N,\M_{\bar{N}})   &=&\frac{\bar{N}}{\bar{N}} = 1.
\end{IEEEeqnarray}
When $\bar{N}=1$ (i.e., only one file or only one user), Inequalities \eqref{eq:finite_set} and \eqref{eq:ratio1} imply that the gap $\xi(K,N,\M)=1$ for all $\M\in[0,N)$, and hence our lower bound is exact.

We therefore assume in the following that $\bar{N}\ge 2$. 
For $\ell \in\{1,\ldots, \bar{N}-1\}$, we have 
\begin{IEEEeqnarray}{rCl}
\Xi(K,N,\M_{\ell}) 
&=& \frac{\frac{N-\frac{N-\ell}{\ell+1}}{\frac{N-\ell}{\ell+1}}\left(1-\left(1-\frac{1}{N}\frac{N-\ell}{\ell+1}\right)^{\bar{N}}\right)}{\sum_{j=1}^{\ell} \left(1-\frac{j}{N-j+1}\frac{N-\ell}{\ell+1}\right)}  \nonumber \\
&=& \frac{\frac{\ell(N+1)}{N-\ell}\left(1-\left(\frac{\ell(N+1)}{(\ell+1)N}\right)^{\bar{N}}\right)}{\sum_{j=1}^{\ell} \left(1+\left(1-\frac{N+1}{N-j+1}\right)\frac{N-\ell}{\ell+1}\right)} \nonumber\\
&=& \frac{\frac{\ell(N+1)}{N-\ell}\left(1-\left(\frac{\ell(N+1)}{(\ell+1)N}\right)^{\bar{N}}\right)}{\frac{\ell(N+1)}{\ell+1} - \frac{(N-\ell)(N+1)}{\ell+1}\sum_{j=1}^{\ell} \frac{1}{N-j+1}} \nonumber\\
&=& \frac{\frac{\ell(\ell+1)}{N-\ell}\left(1-\left(\frac{\ell(N+1)}{(\ell+1)N}\right)^{\bar{N}}\right)}{\ell - (N-\ell)\sum_{j=1}^{\ell} \frac{1}{N-j+1}} \nonumber\\
&\le& \frac{\frac{\ell(\ell+1)}{N-\ell}\left(1-\left(\frac{\ell(N+1)}{(\ell+1)N}\right)^{N}\right)}{\ell - (N-\ell)\sum_{j=1}^{\ell} \frac{1}{N-j+1}} \nonumber\\
&\overset{a=\ell/N}{=}& \frac{\frac{a(\ell+1)}{(1-a)\ell}\left(1-\left(\frac{\ell+a}{\ell+1}\right)^{\ell/a}\right)}{1 - (1-a)\sum_{j=0}^{\ell-1} \frac{1}{\ell-a j}} 
=: \phi(a,\ell). \IEEEeqnarraynumspace
\end{IEEEeqnarray}
Note that since $\ell\in\{1,\ldots, \bar{N}-1\}$,
\begin{equation}
a\in[1/N,1).
\end{equation}
Therefore,
\begin{IEEEeqnarray}{rCl}
\max_{K\in\mathbb{Z}^+}\max_{N\in\mathbb{Z}^+}\max_{\M\in[0,N)} \xi(K,N,\M) 
&\le& \max_{\ell \in \mathbb{Z}^+} \max_{a\in(0,1)} \phi(a,\ell), \IEEEeqnarraynumspace
\end{IEEEeqnarray}
which concludes the proof.

\subsection{Proof of Inequality~\eqref{eq:phipsi}}\label{sec:proofremark}

We have a closer look at the denominator of the function $\phi(a,\ell)$. 

Notice that $\frac{1}{n}\le \int_{n-1}^n\frac{dt}{t}$ for all $n\ge 2$. Therefore, 
\begin{IEEEeqnarray}{rCl}
\sum_{j=0}^{\ell-1} \frac{1}{\ell-a j} &=& \sum_{j=0}^{\ell-1} \frac{1}{\ell(1-a)+a(\ell- j)}\nonumber \\
&\overset{i=\ell-j}{=}& \sum_{i=1}^{\ell} \frac{1}{\ell(1-a)+ai} \nonumber \\
&=& \frac{1}{\ell(1-a)+a} + \frac{1}{a}\sum_{i=2}^{\ell} \frac{1}{\ell(1-a)/a+i} \nonumber\\
&\le& \frac{1}{\ell(1-a)+a} + \frac{1}{a}\sum_{i=2}^{\ell} \int_{\ell(1-a)/a+i-1}^{\ell(1-a)/a+i} \frac{1}{t}\, dt \nonumber\\
&=& \frac{1}{\ell(1-a)+a} + \frac{1}{a}\int_{\ell(1-a)/a+1}^{\ell(1-a)/a+\ell} \frac{1}{t}\, dt \nonumber \\
&=& \frac{1}{\ell(1-a)+a} + \frac{1}{a}\ln\left(\frac{\ell}{\ell(1-a)+a}\right). \label{eq:new_deno}
\end{IEEEeqnarray}

We use \eqref{eq:new_deno} to upper bound the function $\phi(a,\ell)$:
\begin{IEEEeqnarray}{rCl}
\phi(a,\ell) &=& \frac{\frac{a(\ell+1)}{(1-a)\ell}\left(1-\left(\frac{\ell+a}{\ell+1}\right)^{\ell/a}\right)}{1 - (1-a)\sum_{j=0}^{\ell-1} \frac{1}{\ell-a j}} \nonumber \\
&\le& \frac{\frac{a(\ell+1)}{(1-a)\ell}\left(1-\left(\frac{\ell+a}{\ell+1}\right)^{\ell/a}\right)}{1 - \frac{1-a}{\ell(1-a)+a} - \frac{1-a}{a}\ln\left(\frac{\ell}{\ell(1-a)+a}\right)} \nonumber \\
&\overset{b=1/\ell}{=}& \frac{\frac{a(1+b)}{1-a}\left(1-\left(\frac{1+ab}{1+b}\right)^{\frac{1}{ab}}\right)}{1 - \frac{(1-a)y}{1-a+ab} + \frac{1-a}{a}\ln\left(1-a+ab\right)}   \nonumber \\
&= & \psi(a,b). 
\end{IEEEeqnarray}
Noting also that if $\ell > 10^4$, then $b< 10^{-4}$, this concludes the proof of \eqref{eq:phipsi}.

\section{Proof of the Gap-Result in Theorem~\ref{thm:gap_avg}}\label{sec:gap_avg}
We wish to uniformly bound the gap 
\begin{equation}
\theta(K,N,\M) := \frac{\RMavg}{\Rloweravg(\M)},
\end{equation}
irrespective of $K,N\geq 1$ and $\M\in[0,N)$.

Since $ R_{\sf YMA}(K, N, \M)$ upper bounds the rate-memory tradeoff for the worst case, it must also upper bound the rate-memory tradeoff for the average case. Thus, for any number of users $K$ and files $N$:
\begin{IEEEeqnarray}{rCl} 
\label{eq:YMAdec_avg}
\RMavg &\le&R_{\sf YMA}(K, N, \M), \quad \M\in[0,N).
\end{IEEEeqnarray}

We thus have
\begin{IEEEeqnarray}{rCl} \label{eq:new_gap_avg}
\theta(K,N,\M) &\leq& \frac{ R_{\sf YMA}(K, N, \M)}{\Rloweravg(\M)}  \nonumber \\
&\leq & \frac{ R_{\sf YMA}(K, N, \M)}{\RloweravgUnd(K,N,\M)},  \IEEEeqnarraynumspace
\end{IEEEeqnarray}
where we defined 
\begin{IEEEeqnarray}{rCl}\label{eq:deno_avg}
\RloweravgUnd(K,N,\M) &:=& \max_{\ell\in\bar{\mathcal{N}}} \sum_{k=1}^{\ell} \left[\left(1-\frac{1}{N}\right)^{k-1}-\frac{k}{N} \M \right], \IEEEeqnarraynumspace
\end{IEEEeqnarray}
and where the second inequality holds because for all $K,N,\M$: 
\begin{IEEEeqnarray}{rCl}
\Rloweravg(\M) &\geq& \max_{\ell\in\bar{\mathcal{N}}} \left[\Big( 1 - \Big(1- \frac{1}{N}\Big)^\ell \Big)N - \frac{\ell(\ell+1)}{2N}\M \right] \nonumber \\
&=& \max_{\ell\in\bar{\mathcal{N}}} \sum_{k=1}^{\ell} \left[\left(1-\frac{1}{N}\right)^{k-1}-\frac{k}{N} \M \right]. 
\end{IEEEeqnarray}

Define
\begin{IEEEeqnarray}{rCl}
\tilde{\M}_\ell &:=&  \begin{cases}
\frac{N}{\ell+1}\left(1-\frac{1}{N}\right)^{\ell} & \text{ if } \ell\in\{0,1,\ldots,\bar{N}-1\}, \\
 0 & \text{ if } \ell = \bar{N}.
\end{cases}
\end{IEEEeqnarray}
and note that $0= \tilde{\M}_{\bar{N}} < \tilde{\M}_{\bar{N}-1} < \cdots < \tilde{\M}_0 = N$.
Using similar arguments as in Appendix~\ref{app:piecewise}, it can be shown that function  $\RloweravgUnd(K,N,\M)$  is piecewise-linear with $\bar{N}$ line segments over the intervals
\begin{subequations}\label{eq:intervals_avg}
\begin{IEEEeqnarray}{rCl}
\ [\tilde{\M}_{\ell}, \ \tilde{\M}_{\ell-1}], & &\qquad \ell \in\{2, \ldots, \bar{N}\}\\
\ [\tilde{\M}_1, \ \tilde{\M}_0).&&
\end{IEEEeqnarray}
\end{subequations}

We next upper bound $R_{\sf YMA}(K,N,\M)$ by a function $\overline{R}_{\sf YMA}(K,N,\M)$ that is piecewise-linear over the same intervals  \eqref{eq:intervals_avg}. 
Specifically, for every $\ell\in\bar{\mathcal{N}}$, define for $\M \in [\tilde{\M}_\ell, \tilde{\M}_{\ell-1})$:
\begin{IEEEeqnarray}{rCl}
\overline{R}_{\sf YMA}(K,N,\M) &:=&{  \frac{\tilde{\M}_{\ell-1}-\M}{\tilde{\M}_{\ell-1}-\tilde{\M}_{\ell}} }  R_{\sf YMA}(K,N,\tilde{\M}_{\ell})\nonumber \\
 &&  + { \frac{ \M- \tilde{\M}_\ell}{\tilde{\M}_{\ell-1}-\tilde{\M}_\ell} } R_{\sf YMA}(K,N,\tilde{\M}_{\ell-1}). \IEEEeqnarraynumspace
\end{IEEEeqnarray}
Notice  that 
\[\overline{R}_{\sf YMA}(K,N,\tilde{\M}_\ell)
= R_{\sf YMA}(K,N,\tilde{\M}_\ell), \quad \forall \ell\in \{1,\ldots,\bar{N}\}, \] whereas for general $\M\in[0,N)$:
\begin{equation}\label{eq:new_upper_avg}
\overline{R}_{\sf YMA}(K,N,\M) \geq {R}_{\sf YMA}(K,N,\M),
\end{equation}
because $R_{\sf YMA}(K,N,\M)$ is convex {in $\M$ for fixed $K,N$.} 

Plugging \eqref{eq:new_upper_avg} into \eqref{eq:new_gap_avg}, we obtain: 
\begin{equation}\label{eq:newest_gap_avg}
\theta(K,N,\M) \leq \frac{\overline{R}_{\sf YMA}(K,N,\M) }{ \RloweravgUnd(K,N,\M)} =: \Theta(K,N,\M).
\end{equation} 
Following similar arguments as in the proof of Theorem~\ref{thm:gap_worst}, we have 
\begin{IEEEeqnarray}{rCl}\label{eq:bound2}
\max_{\M\in[0,N)} \theta(K,N,\M) & \leq & \max_{\ell\in\bar{\mathcal{N}}}  \Theta(K,N,\tilde{\M}_\ell). 
\end{IEEEeqnarray}

{We continue to bound the right-hand side of \eqref{eq:bound2}.} Irrespective of $K,N \in \mathbb{Z}^+$, we have:
\begin{IEEEeqnarray}{rCl}
\Theta(K,N,\tilde{\M}_{\bar{N}})  &=& \frac{\bar{N}}{N(1-(1-1/N)^{\bar{N}})} \nonumber\\
&=& \frac{x}{1-((1-1/N)^N)^x} \Big|_{x=\bar{N}/N} \nonumber \\
&\overset{(a)}{\le}& \frac{x}{1-e^{-x}} \Big|_{x=\bar{N}/N} \nonumber \\
&\overset{(b)}{\le}& \frac{1}{1-e^{-1}}, 
\end{IEEEeqnarray}
where $(a)$ follows since $(1-1/\zeta)^\zeta \le e^{-1}$ for all $\zeta > 1$ and $(b)$ follows since $x\mapsto\frac{x}{1-e^{-x}}$ is an increasing function. 
This implies that when $\bar{N}=1$ (i.e., one file or one user), then $\Theta(K,N,\M) \le \frac{1}{1-e^{-1}} \leq 1.582$ for all $\M\in[0,N)$. 

In the following, we assume that $\bar{N}\ge 2$. 
As for $\ell \in\{1,\ldots,\bar{N}-1\}$, we have 
\begin{IEEEeqnarray}{ll}
& \Theta(K,N,\M_\ell) \nonumber\\
&=\frac{\frac{N-\frac{N}{\ell+1}\left(1-\frac{1}{N}\right)^{\ell}}{\frac{N}{\ell+1}\left(1-\frac{1}{N}\right)^{\ell}}\left(1-\left(1-\frac{1}{N}\frac{N}{\ell+1}\left(1-\frac{1}{N}\right)^{\ell}\right)^{\bar{N}}\right)}{N-\left(N+\frac{\ell}{2}\right)\left(1-\frac{1}{N}\right)^{\ell}} \nonumber\\
&= \frac{\left(\frac{\ell+1}{N}-\frac{1}{N}\left(1-\frac{1}{N}\right)^{\ell}\right)\left(1-\left(1-\frac{1}{\ell+1}\left(1-\frac{1}{N}\right)^{\ell}\right)^{\bar{N}}\right)}{\left(1-\frac{1}{N}\right)^{\ell}\left(1-\left(1+\frac{\ell}{2N}\right)\left(1-\frac{1}{N}\right)^{\ell}\right)} \nonumber \\
&\le \frac{\left(\frac{\ell+1}{N}-\frac{1}{N}\left(1-\frac{1}{N}\right)^{\ell}\right)\left(1-\left(1-\frac{1}{\ell+1}\left(1-\frac{1}{N}\right)^{\ell}\right)^{N}\right)}{\left(1-\frac{1}{N}\right)^{\ell}\left(1-\left(1+\frac{\ell}{2N}\right)\left(1-\frac{1}{N}\right)^{\ell}\right)} \nonumber \\
&\overset{(a)}{=} \frac{\left(u+v-v\left(1-v\right)^{u/v}\right)\left(1-\left(1-\frac{v}{u+v}\left(1-v\right)^{u/v}\right)^{1/v}\right)}{\left(1-v\right)^{u/v}\left(1-\left(1+\frac{u}{2}\right)\left(1-v\right)^{u/v}\right)} \nonumber\\
&=: \eta(u,v), 
\end{IEEEeqnarray}
where $(a)$ follows by a change of variable $u=\ell/N$ and $v=1/N$. 
Note that since $\ell \in\{1,\ldots,\bar{N}-1\}$ and assuming $\bar{N}\ge 2$, 
\begin{IEEEeqnarray}{rCl}
u \in \left[\frac{1}{N},\frac{\bar{N}-1}{N}\right] & \text{ and } & v \in (0,1/2]. 
\end{IEEEeqnarray}
Also, it holds that $\frac{1}{1-e^{-1}} < \max_{u\in(0,1]}\max_{v\in(0,1/2]} \eta(u,v)$. Therefore,
\begin{IEEEeqnarray}{rCl}
\max_{K\in\mathbb{Z}^+}\max_{N\in\mathbb{Z}^+}\max_{\M\in[0,N)} \theta(K,N,\M) 
&\le& \max_{u\in(0,1]}\max_{v\in(0,1/2]} \eta(u,v),\nonumber \\
\end{IEEEeqnarray}
which concludes the proof.

{\section{Conclusion}
This paper derives new lower bounds on the rate-memory tradeoff under a worst-case or an average-case scenario. The obtained lower bounds are compared to upper bounds on the rate-memory tradeoffs in decentralized caching scenarios and shown to match these upper bounds up to a multiplicative gap of at most $2.315$ (in the worst-case scenario) and $2.507$ (in the average-case scenario). Previous bounds could establish a gap of $4.7$ and $4$, respectively. In a work that is parallel to this \cite{YMA:17}, improved upper bounds were presented which match the decentralized upper bounds up to factors of almost $2$. The bounds in \cite{YMA:17} are based on similar  technical steps as used in this paper. The improvement is obtained through an additional averaging step over the labeling of the receivers.

The converse technique presented in this paper can be extended  to setups where delivery communication takes place over a noisy broadcast channel  (BC) (rather than a noise-free link as considered in this paper). Corresponding bounds on the capacity-memory tradeoff over general discrete memoryless  BCs can be found in  \cite{SaeediWiggerYener-2016}.

}

%

\appendices

	\section{Proof of Equation \eqref{eq:equiv1}}\label{app:piecewise}

	Fix  $i \in \bar{\set{N}}$ and $\M \in [\M_{i}, \M_{i-1}]$. 
Define 
\begin{align}
\nu_j:=\frac{1}{M_{j-1}}=\frac{k}{N-j+1}, \qquad j\in\{1,\ldots, \bar{N}\}.
\end{align}
Notice that 
\begin{equation}
\label{eq:nu_ordering}
\nu_1\leq\nu_2\leq\ldots\leq\nu_{\bar{N}}
\end{equation} and  
\begin{equation}\label{eq:sandwich}
\frac{1}{\nu_{i+1}} \leq  M  \leq \frac{1}{\nu_{i}}.
\end{equation}
Rewrite  $\RlowerworstUnd(K,N,\M)$ as 
	\begin{IEEEeqnarray}{rCl}\label{eq:max}
\RlowerworstUnd(K,N,\M)
		& = & 
		\max_{\ell \in \bar{\set{N}}} \Bigg[ \sum_{j=1}^\ell \big(1 - \M \cdot   \nu_j \big)\Bigg].
	\end{IEEEeqnarray}
By \eqref{eq:nu_ordering} and \eqref{eq:sandwich}, the summands $(1- M \cdot \nu_j)$ are positive or zero for all $j \leq  i$ and they are negative  $j> i$. The maximum in \eqref{eq:max} is thus achieved by choosing  $\ell=i$. This proves equation~\eqref{eq:equiv1}.

	\section{Proof of Lemma~\ref{lem:alphas}}\label{app:lem3}
	We first prove that for each $k\in\{1,\ldots, \ell\}$:
	\begin{equation}\label{eq:whattoprove1}
	{\alpha}_{k}  \leq \frac{k \M}{N-k+1},
	\end{equation}
	which establishes the upper bound
	\begin{equation}\label{eq:prove1}
	\sum_{k=1}^\ell {\alpha}_{k}  \leq   \sum_{k=1}^\ell \frac{k \M}{N-k+1}.
	\end{equation}
	
	
	For each partial demand vector $\tilde{\d}=(d_1, \ldots, d_{k-1})$, let $W_{\tilde{\d}}:=\{W_{ d_1}, \ldots, W_{d_{k-1}}\}$. We have:
	\begin{align}
	&F\alpha_k\nonumber \\
	&=\frac{1}{\ell!{N \choose \ell}}\sum_{\mathbf{d}\in\Qell}I(W_{d_k};\V_1,\ldots, \V_k|W_{d_1},\ldots,W_{d_{k-1}})\nonumber\\
	&=\frac{1}{\ell!{N \choose \ell}} \sum_{\tilde{\d}\in \Qkone} \sum_{\substack{\mathbf{d}\in\Qell \colon \\ (d_1,\ldots, d_{k-1})=\tilde{\d}}} I(W_{d_k};\V_1,\ldots, \V_k|W_{\tilde{\d}})\nonumber\\
	&\stackrel{(a)}{=}\frac{1}{\ell!{N \choose \ell}} \sum_{\tilde{\d}\in \Qkone} \; \sum_{j \in\set{N}\backslash \tilde{\d}} I(W_{j};\V_1,\ldots, \V_k|W_{\tilde{\d}}) \nonumber \\
	&\hspace{3.4cm} \cdot {{N-k} \choose {\ell-k}} (\ell-k)!\nonumber\\
	&=\frac{1}{{k!}{N\choose k}} \sum_{\tilde{\d}\in \Qkone} \; \sum_{j \in\set{N}\backslash \tilde{\d}}  I(W_{j};\V_1,\ldots, \V_k |W_{\tilde{\d}})\nonumber\\
	&\stackrel{(b)}{=}\frac{1}{{k!}{N\choose k}}  \bigg[\sum_{\tilde{\d}\in \Qkone} \Big[{ H\big(\{W_j \colon j \in\set{N}\backslash \tilde{\d}\}\,  \big|\, W_{\tilde{\d}}\big) }- \nonumber \\
	& \hspace{3cm} \sum_{j \in\set{N}\backslash \tilde{\d}}  H(W_{j}|\V_1,\ldots, \V_k , W_{\tilde{\d}}) \Big]\bigg]\nonumber\\
	&\stackrel{(c)}{\leq}\frac{1}{{k!}{N \choose k}}\sum_{\tilde{\d}\in \Qkone}{ I\big(\{W_j \colon j \in\set{N}\backslash \tilde{\d}\}\, ;\, \V_1,\ldots, \V_k \big | W _{\tilde{\d}}\big)} \nonumber\\
	&\stackrel{(d)}{\leq} \frac{(k-1)!{N \choose k-1}}{{k!}{N \choose k}}k F\M \nonumber\\
	&={\frac{ kF\M}{N-k+1}},
	\end{align}
	where $(a)$ holds because for each value of $\ell$ and $j$ there are ${{N-k} \choose {\ell-k}} (\ell-k)!$ ordered demand vectors $\d=(d_1,\ldots, d_K)\ \in \Qell$ with $(d_1,\ldots, d_{k-1})=\tilde{\d}$ and with $d_k=j$;  (b) holds  by the independence of the messages; (c) holds because for any random tuple $(A_1, \ldots, A_L)$ it holds that $\sum_{l=1}^L H(A_l) \geq H(A_1, \ldots, A_L)$; and (d) holds because $ I(W_1,\ldots, W_N;\V_1,\ldots, \V_k|W _{\tilde{\d}})$ cannot exceed $kF\M$. This concludes the proof of \eqref{eq:whattoprove1} and thus of \eqref{eq:prove1}.

	We now prove 
	\begin{equation}
	\sum_{k=1}^\ell {\alpha}_{k}  \leq  \frac{\ell^2 \M}{N}.
	\end{equation}
	For each  $\d\in \Qell$: 
	\begin{IEEEeqnarray}{rCl}
		\lefteqn{I(W_{d_1}; \V_1)+ \sum_{k=2}^\ell I(W_{d_k}; \V_1, \ldots, \V_k|W_{d_1}, W_{d_2}, \ldots, W_{d_{k-1}})} \quad \nonumber \\
		&  \leq & I(W_{d_1},W_{d_2}, \ldots, W_{d_{\ell}}; \V_1, \ldots, \V_\ell). \hspace{2.3cm}
	\end{IEEEeqnarray}
	So,
	\begin{align}
	&F  { N\choose \ell} {\ell!} \cdot\sum_{k=1}^\ell {\alpha}_{k} \nonumber \\
	&=\sum_{\d\in\Qell} \bigg[ I(W_{d_1}; \V_1) \nonumber \\
	& \qquad \qquad \;+\sum_{k=2}^\ell I(W_{d_{k}}; \V_1, \ldots, \V_k|W_{d_1}, W_{d_2}, \ldots, W_{d_{k-1}}) \bigg]  \qquad \nonumber \\
	&\quad  \leq   \sum_{\d \in \Qell} I(W_{d_1}, W_{d_2},\ldots, W_{d_{\ell}}; \V_1 \ldots, \V_\ell) \nonumber \\
	&\quad  =  {\ell!}  \sum_{\substack{\d \in \Qell\colon\\ d_1<d_2\cdots<d_\ell}} I(W_{d_1}, W_{d_2},\ldots, W_{d_{\ell}}; \V_1 \ldots, \V_\ell) \nonumber \\
	&\quad  \stackrel{(a)}{\leq}  {\ell!} {N \choose \ell} \frac{\ell}{N} I( W_1, \ldots, W_N; \V_1, \ldots, \V_\ell)\nonumber\\
	&\quad \leq    \frac{\ell}{N}{\ell!} { N\choose \ell} \ell F\M,
	\end{align}
	where $(a)$ follows by Lemma~\ref{prop:han}.
	\vspace{1mm}

\section{Proof of Lemma~\ref{lem:betas}}\label{app:lem11}
	We first prove that for each $k\in\{1,\ldots, \ell\}$:
	\begin{equation}\label{eq:whattoprove3}
	{\beta}_{k}  \leq \frac{k \M}{N},
	\end{equation}
	which establishes the upper bound
	\begin{equation}\label{eq:prove3}
	\sum_{k=1}^\ell {\beta}_{k}  \leq   \sum_{k=1}^\ell \frac{k \M}{N}.
	\end{equation}
	Defining $\D_{k}:=(D_1, \ldots, D_{k})$, we have:
	\begin{IEEEeqnarray}{rCl}
		F\beta_k
		&=& I(W_{D_k};\V_1,\ldots, \V_k|W_{D_1},\ldots,W_{D_{k-1}},\D) \nonumber\\
		&=& I(W_{D_k};\V_1,\ldots, \V_k|W_{D_1},\ldots,W_{D_{k-1}},\D_k) \nonumber\\
		&\stackrel{(a)}{=}& \frac{1}{ N^k }  \sum_{\substack{\mathbf{d}\in\Qrep_k}}I(W_{d_k};\V_1,\ldots, \V_k|W_{d_1},\ldots,W_{d_{k-1}})\nonumber\\
		&=& \frac{1}{ N^k }  \sum_{\substack{\tilde{\mathbf{d}}\in\Qrep_{k-1}}} \; \sum_{j=1}^N I( W_j ;\V_1,\ldots, \V_k|{W}_{\tilde{\mathbf{d}}} ) \nonumber \\
		&\stackrel{(b)}{\leq}& \frac{1}{ N^k }  \sum_{\substack{\tilde{\mathbf{d}}\in\Qrep_{k-1}}} \; I( W_1,\ldots, W_N ;\V_1,\ldots, \V_k|{W}_{\tilde{\mathbf{d}}} ) \nonumber \\
		&\stackrel{(c)}{\leq}& \frac{1}{ N^k }  \sum_{\substack{\tilde{\mathbf{d}}\in\Qrep_{k-1}}} k F\M \nonumber\\
		&=& \frac{kF\M}{N},
	\end{IEEEeqnarray}
	where $(a)$ holds by writing out the conditioning on $\mathbf{D}_k$ in form of  an expectation; $(b)$ holds because the messages are independent and because $H(A_1,\ldots, A_L) \leq \sum_{l=1}^L H(A_l)$ for any random $L$-tuple $(A_1,\ldots, A_L)$; and  (c) holds  because $I(W_{1}, \ldots, W_N;\V_1,\ldots, \V_k|{W}_{\tilde{\mathbf{d}}})$ cannot exceed $kF\M$. This concludes the proof of \eqref{eq:whattoprove3} and thus \eqref{eq:prove3}.

	We now prove 
	\begin{equation}
	\sum_{k=1}^\ell {\beta}_{k}  \leq  \E_\D\big[\kappa_\D(\ell)\big]\cdot \frac{\ell \M}{N}.
	\end{equation}
	
	Let $\D^{\textnormal{dist}}_\ell$ be a vector containing all distinct elements of $\D_\ell:=(D_1,\ldots, D_\ell)$. 
	Notice that $\D^{\textnormal{dist}}_\ell$ is of length $\kappa_{\D_\ell}(\ell)$. Also, following the definition of the previous section,  $W_{\D_\ell}:=\{W_{D_1},\ldots, W_{D_\ell}\}=W_{\D^{\textnormal{dist}}_\ell}$. We have:
	\begin{IEEEeqnarray}{ll}
		&F \sum_{k=1}^\ell {\beta}_{k}  \nonumber\\
		&= I(W_{D_1}; \V_1 |\D) \nonumber \\
		& \hspace{0.5cm} +\sum_{k=2}^\ell I(W_{D_k}; \V_1, \ldots, \V_k|W_{D_1}, W_{D_2}, \ldots, W_{D_{k-1}},\D) \nonumber \\
		&\le  I(W_{\D_\ell}; \V_1, \ldots, \V_\ell|\D) \nonumber\\
		&\le  I(W_{\D_\ell}; \V_1, \ldots, \V_\ell|\D_{\ell}) \nonumber\\
		&=  I\left(W_{\D_\ell}; \V_1, \ldots, \V_\ell \big|\D_\ell, \kappa_{\D_\ell}(\ell)\right) \nonumber \\
		&=  \sum_{i=1}^\ell \mathbb{P}(\kappa_{\D_\ell}(\ell)=i) I\left(W_{\D_\ell} ; \V_1, \ldots, \V_\ell \big|\D_\ell,\kappa_{\D_\ell}(\ell)=i\right) \nonumber \\
		&\overset{(a)}{=}  \sum_{i=1}^\ell \mathbb{P}(\kappa_{\D_\ell}(\ell)=i) I\left(W_{\D_\ell^{\textnormal{dist}}} ; \V_1, \ldots, \V_\ell \big|\D_\ell^{\textnormal{dist}}, \kappa_{\D_\ell}(\ell)=i\right) \nonumber \\
		&\overset{(b)}{=}  \sum_{i=1}^\ell \mathbb{P}(\kappa_{\D_\ell}(\ell)=i) \sum_{\substack{\tilde{\d} \in \mathcal{Q}_i^{\textnormal{dist}}\colon \\\tilde{d}_1 < \tilde{d}_2 \cdots < \tilde{d}_i}} \frac{1}{{N \choose i}}I(W_{\tilde{\d}}; \V_1, \ldots, \V_\ell)  \nonumber\\
		&\overset{(c)}{\leq} \sum_{i=1}^\ell \mathbb{P}(\kappa_\D(\ell)=i) \frac{i}{N} I(W_1,\ldots,W_N; \V_1, \ldots, \V_\ell) \nonumber \\
		&\overset{(d)}{\le} \sum_{i=1}^\ell \mathbb{P}(\kappa_\D(\ell)=i)\cdot i \cdot \frac{\ell F\M}{N} \nonumber \\
		&= \E_\D\big[\kappa_\D(\ell)\big] \cdot \frac{\ell F\M}{N}.
	\end{IEEEeqnarray}
Notice that here
	(a) holds {because for the involved mutual informations only the set of distinct demands matters and not the exact demand vector.} 
 		Moreover, (b) holds because given $\kappa_{\D_\ell}(\ell)=i$ the probability that $\D_\ell^{\textnormal{dist}}$ equals a specific vector $\tilde{\d}\in\set{Q}_i^{\textnormal{dist}}$ equals ${{N \choose i}}^{-1}$; (c) follows from Lemma~\ref{prop:han}; and (d) follows since $ I(W_{1}, \ldots, W_{N};  \V_1, \ldots, \V_\ell) $ cannot be larger than $\ell F\M$.

\bibliographystyle{IEEEtran}
\bibliography{IEEEabrv,bibfile}

\begin{IEEEbiographynophoto}{Chien-Yi Wang} received the B.S. degree in electrical engineering from
National Tsing Hua University, Hsinchu, Taiwan, in 2007, the M.S. degree
in electronics engineering from National Taiwan University, Taipei, Taiwan,
in 2010, and the PhD degree in Computer and Communication Sciences
from the \'Ecole Polytechnique F\'ed\'erale (EPFL), Lausanne, Switzerland,
in 2015. During the academic year 2009-2010, he was an exchange student at
RWTH Aachen University, Aachen, Germany. In 2016, he was a postdoctoral
researcher in the Communications and Electronics Department at Telecom
ParisTech. Currently he is a senior engineer at
MediaTek. His research interests include network information theory and
wireless communications.
\end{IEEEbiographynophoto}

\begin{IEEEbiographynophoto}{Shirin Saeedi Bidokhti}(S'09-M'12)
is a research assistant professor in the Department of Electrical and
Systems Engineering at the University of Pennsylvania (UPenn). She
received the B.Sc. degree in Electrical Engineering from University of
Tehran in 2005, and the M.Sc and Ph.D. degrees in Communication
Systems from \'Ecole Polytechnique F\'ed\'erale de Lausanne (EPFL) in
2007 and 2012, respectively. Before joining UPenn in 2017, she was a
postdoctoral fellow with the Institute for Communications Engineering
at the Technische Universit\"at M\"unchen (2013-2015), a postdoctoral
fellow with the Department of Electrical  Engineering at Stanford
University (2015-2017), and a postdoctoral research scholar with the
Department of Electrical Engineering at the Pennsylvania State
University (2017).  She has been awarded an  Advanced Postdoc Mobility
Fellowship (2014) and a Prospective Researcher Fellowship (2012) both
from the Swiss National Science Foundation. Her research interests
include network information theory, network coding, and data
compression.\end{IEEEbiographynophoto}

\begin{IEEEbiographynophoto}
{Mich\`ele Wigger} (S'05, M'09, SM'14) received the M.Sc. degree in electrical engineering, with distinction, and the Ph.D. degree in electrical engineering both from ETH Zurich in 2003 and 2008, respectively. In 2009, she was first a post-doctoral fellow at the University of California, San Diego, USA, and then joined Telecom Paris Tech, Paris, France, where she is currently a Full Professor. Dr. Wigger has held visiting professor appointments at the Technion-Israel Institute of Technology and ETH Zurich. Dr. Wigger has previously served as an Associate Editor of the IEEE Communication Letters, and is now Associate Editor for Shannon Theory of the IEEE Transactions on Information Theory. She is currently also serving on the Board of Governors of the IEEE Information Theory Society. Dr. Wigger's research interests are in multi-terminal information theory, in particular in distributed source coding,  capacities of networks, and distributed hypothesis testing.
\end{IEEEbiographynophoto}

\end{document}